\documentclass[manuscript]{acmart}
\usepackage{amsmath}
\usepackage[most]{tcolorbox}
\usepackage{enumitem}

\AtBeginDocument{%
  }

\begin{document}

\title[A Model of Integrated Information Processing in Human-AI Interaction]{A Model of Integrated Information Processing in Human-AI Interaction}



\author{Tim Schrills}
\email{tim.schrills@uni-luebeck.de}
\orcid{000-0001-7685-1598}
\affiliation{%
  \institution{University of Luebeck}
  \city{Luebeck}
  \country{Germany}
}

\author{Thomas Franke}
\email{thomas.franke@uni-luebeck.de}
\orcid{0000-0002-7211-3771}
\affiliation{%
  \institution{University of Luebeck}
  \city{Luebeck}
  \country{Germany}
}

\renewcommand{\shortauthors}{Schrills \& Franke}


\begin{abstract}
For the field of Human--AI Interaction (HAII) research to move forward, theoretical work that links psychological mechanisms to interface design is needed. While examining theoretical models, research should extend rather than replace established HCI and automation research, and thus, adapt to increasing autonomy and agency of AI systems. Building on prior frameworks that focus on roles and levels in human interaction with automation, an existing gap remains from a psychological view: a task-centred, process-oriented account that links psychological mechanisms of action regulation to concrete design and evaluation levers for human–AI coupling, expressed in a unified vocabulary for human and machine. On top of that, existing models may help to describe how a system is designed (e.g., models of function allocation in automation) but fall short in demonstrating how this design affects human behavior. We present the Integrated Information Processing (IIP) model, a task-centered, cybernetic model that conceptualizes humans, machines, and their joint activity as coupled control loops. The IIP model uses a unified modeling language for human and artificial agents, making psychological models of action regulation accessible for AI system design. 

As a core feature of the IIP model, we argue that efficacy within a shared task is characterized by three integration qualities---\textit{input adequacy}, \textit{reference consonance}, and \textit{output operativity}---which critically influence benchmarks of human-centeredness like  \textit{transparency} and \textit{controllability}. The IIP model provides practical criteria that map interface choices (e.g., XAI techniques) to theory-driven expectations of user behavior, offering guidance for designing and evaluating user interfaces. To this end, we present (1) a continuity-preserving theoretical discourse that extends HAII to agency in AI; (2) the IIP model with three information-processing qualities; and (3) applications of the IIP model for exemplary use cases demonstrating implications for interface design.
\end{abstract}



\maketitle

\section{Introduction}

Imagine an agent with a single task: to write the first sentence for a scientific article. This process can be described as a cybernetic control loop: The agent begins with information it possesses about the article's content and its main messages (the input function), along with its goals—to draw attention, signal the scientific discipline, or set a particular tone (the reference function). Based on this information, the agent chooses words to fill the blank page (the output function). The sentence it produces then becomes a new piece of information that the agent evaluates. The interesting part of this scenario is that this agent could be a human or a machine. In this example, we utilized a cybernetic model to describe the process, which is fit to structure the information processing and resulting actions of both human (or animal) and machine \cite{Wiener1948Cybernetics}. In fact, several key psychological models on human action regulation are based on control theoretic conceptions (e.g., \cite{LazarusFolkman1984,carverStructureBehavioralSelfRegulation2000, fuller_towards_2005}). At the same time, cybernetics is the fundamental architecture in control engineering and is also currently proposed as a key conceptualization of artificial intelligence (AI) systems \cite{floridi_ai_2025}. The fact that control loops are applicable in both cases makes them useful in situations where both humans and AI are involved in tasks. 

Models of shared notion for human and AI are needed, due to an increasing number of everyday tasks in professional (e.g., \cite{blease_generative_2024}) and non-professional (e.g., \cite{SimonNielsenFletcher2025_GenAI_News}) contexts that are not carried out by humans alone. That is, professions with varying levels of expertise, as well as varying fields, are facing automation of information processing, especially through the means of augmentation through artificial intelligence (AI). In response to AI's effects on human activity, industrial guidelines \cite{GooglePAIR_Guidebook_2019}, ethical considerations \cite{flathmann2021modeling}, as well as political regulations \cite{eu_ai_act_2024} require AI systems to be 'human-centered' - a term with diverging operationalization based on the research discipline discussing it. Parallel to this, recent research is driven by technological solutions targeted to enable human-centered AI systems, most prominently illustrated through the field of explainable AI (XAI, see \cite{gunningDARPAExplainableArtificial2019a}). XAI techniques, along with similar approaches such as human-aware AI \cite{sreedharanHumanawareAIFoundational2023}, may allow users to better understand AI systems and, in theory, improve users' ability to effectively control the system (i.e., one key regulatory goal of effective human oversight). That is, by improving the understandability and controllability of AI systems, designers want to enable relationships between human and AI that create the characteristics of teaming \cite{attigMoreTaskPerformance2024}, cooperation, or other forms of synergistic (like described in \cite{rieger_explaining_2025}) interaction.

What seems to be the fundamental driving force behind the described approaches and regulations is to identify, enable, and optimize how human and AI-driven information processing and action regulation can be best coupled; that is, benefiting from machine and human capabilities for information processing at the same time. In this case, 'coupling' refers to any method that allows the exchange of signals through which humans and machines can engage in shared information processing. 
In accordance with previous research in human-machine interaction \cite{parasuramanModelTypesLevels2000a}, such a coupling includes systems where human and machine agents may exhibit varying levels of autonomy and control (cf.\cite{shneidermanHumanCenteredArtificialIntelligence2020a}). However, to not only enable humans and machines to exchange signals but to optimize their shared information processing, research may focus on Human-AI \textit{integration}, i.e., the optimal coupling of human and machine information processing.

This paper proposes the integrated information processing (IIP) model—a compact and applicable approach to modeling Human-AI integration by adapting established theories of control in human-machine systems to the unique challenges of high-autonomy systems. A primary strength of this model is its use of a single, consistent structure—the cybernetic control loop—to describe both human and AI information processing and action regulation. The present paper explicates how such cybernetic control loops can be adopted to model shared information processing, and it defines key integration qualities—input adequacy, reference consonance, and output operativity—to guide the design and evaluation of resulting systems combining human and artificial agency.

To achieve this, the present paper first explicates the key aspects of cybernetic control theory and its relationship to human-machine interaction and automation research, serving as the theoretical foundation (section 2). It then introduces the IIP Model (section 3), which details the mechanisms of coupling and shared information processing as well as integration qualities as part of the IIP model (section 4). Following this, model applications fields (section 5) are explicated, providing practical examples of how the IIP model can be used to describe, design, and evaluate human-AI systems. Finally, the discussion and conclusion  (sections 6-7) address the key paths for the further development of the IIP model, addressing its current limitations and present resulting research questions.

\section{Theoretical Foundations}

\subsection{AI and Automation: Continuity and Design Questions}
Given that AI systems introduce new system dynamics and capabilities, they may appear like a distinct technology to users, although the structural psychological dynamics of HAII are similar to previous technologies. That is, while AI systems introduce new technological capabilities, especially to deal with problems without a previously determined solution, the underlying structure of how they affect users remains similar to technologies previously examined in research on human-automation interaction. For example, automating a heating system by using a neural network may challenge the user's situation awareness on when and how a heater turns on or off, which is a challenge known from automation research, only intensified (i.e., not structurally changed) through the opaque nature of the method. 

The key point is that contemporary AI systems call to extend, but do not invalidate, decades of research on automation. Systems studied in automation research typically embed fixed rules and narrow feedback loops (i.e., using fewer variables) engineered by humans for well-specified tasks, with limited adaptability and agency \cite{parasuraman_situation_2008}. By contrast, present AI systems acquire their algorithmic processing from data (i.e., generate models which are utilized in tasks), dynamically update their control policies (i.e., to adapt to different users), and generalize across task variations due to representation learning \cite{GoodfellowBengioCourville2016}. Still, the design questions automation research has raised over the past decades remain central (and partially unsolved): (1) what functions are to automate (i.e., the quest of function allocation \cite{Sheridan2000_FunctionAllocation}), (2) to what degree should these functions be automated (i.e., finding optimal characteristics of automation \cite{EndsleyKaber1999_LoA, onnaschHumanPerformanceConsequences2014b}), and (3) with what consequences for human performance (e.g. ironies of automation, \cite{bainbridgeIroniesAutomation1983}) and the human capability to take responsibility \cite{floridi_morality_2004}. Framing AI as a continuation of automation, and therefore positioning human-AI interaction research in reference to human-automation interaction research, clarifies both: where it is helpful to continue existing research paradigms (e.g., feedback, allocation of control, error management) and where to address challenges tied to new technological advances (e.g., trained rather than engineered models, non-deterministic outputs, emergent behavior). 

\subsection{A Cybernetic View: Nested Control and Implications for HCI}
A cybernetic (i.e., control theoretic) view makes this continuity of AI as a further form of automation explicit. Sheridan’s definition of automation as a system that \emph{senses}, \emph{processes/decides}, and \emph{acts} according to rules (or learned policies) foregrounds self-control as the property that binds these functions into a closed loop with a human \cite{sheridanTeleroboticsAutomationHuman1992}. In AI, the same self-control is increasingly distributed across software components—planners, learners, monitors—so the system’s effective “agency” grows. Floridi and Sanders’ account of artificial agents as entities assessable at the level of their agency (i.e., interactivity, autonomy, and adaptivity), provides a useful lens for this growth of agency \cite{floridi_morality_2004}. 

The cybernetic control loop offers a compact description that can be useful to describe AI systems exhibiting high levels of agency. In Wiener’s original formulation, a controller compares a \emph{perceived} state with a \emph{reference} and closes the loop by acting (i.e., generating an \textit{output}) to reduce discrepancy under noise and delay \cite{Wiener1948Cybernetics}. Control-theoretic psychology translated this scheme to human behavior: Powers’ perceptual control theory treated behavior as control of perception, and Carver \& Scheier elaborated multi-level, goal-referenced loops that regulate action and even affect \cite{Powers1973_BehaviorControlPerception, CarverScheier1982_ControlTheory}. For human-computer interaction, two implications from these psychological models matter: First, an input “function” is not a scalar value but a pipeline of transformations—selection, abstraction, aggregation, and uncertainty management—that determine what \emph{is perceived} as the state to be controlled; modern AI intensifies this because learned representations of AI systems act as adaptive input functions, which are more complicated to control and deal with for human users. Second, loops are applicable for explaining both cognitive information processing within action regulation and motivational dynamics (e.g., affect as a signal of control error), which shape user behavior around AI systems. That is, the combination of input, reference, and output enables researchers to describe how an agent behaves, yet also what triggers its behavior. 

Empirically, the human–automation literature shows why conceptualizing human behavior from this perspective of control loops matters. Levels of automation (see \cite{EndsleyKaber1999_LoA}) reallocate sensing, deciding, and acting across human and machine, with well-documented trade-offs, including the out-of-the-loop performance problem and automation-induced complacency—phenomena that reappear with intelligent systems \cite{parasuramanModelTypesLevels2000a, ParasuramanMolloySingh1993, EndsleyKiris1995}. AI intensifies such core risks of miscalibrated trust, loss of situation awareness, and skill degradation. 

The input function’s growing complexity is pivotal to motivating why automation research faces additional challenges with the rise of AI systems. Deep learning, for example, replaces much human engineering of how to combine raw data purposefully with learned representations that map raw signals to abstract states suited for control and prediction \cite{GoodfellowBengioCourville2016}. As a result, the “perceived” state is itself the product of nested estimation loops of an AI agent (e.g., through filtering, outlier rejection, uncertainty estimation), which may shape system behavior as much as the decision-making process that will follow. Treating inputs as functions rather than values thus aligns cybernetics with modern forms of machine learning, especially when considering that these functions are composed of other functions that, together, form into the input function related to the task at hand. 

There is no single breakpoint that turns complex self-control into “agency”, yet nested and interacting loops plausibly yield higher degrees of agent-like behavior. For HCI, the design challenge is not whether to label such systems “agents,” but how to anticipate the user-facing consequences of deeper nesting—e.g., longer chains of internal state that are opaque to users or more variable and unforeseeable emergence of system behavior. These challenges call to extend models of human–AI interaction from a cybernetic perspective of nested control loops to make the structure of self-control explicit. 

Concrete examples illustrate how nested loops arise across the three control functions—input, reference, and output. Revisiting the example from the beginning of this paper, in the (1) \emph{input} function, an LLM-based writing assistant might start with a prompt as input, but include an online search or access to otherwise stored data to effectively fulfill a request. That is, input information is gathered utilizing additional functions-e.g., to determine at what point an online search as part of the information gathering was sufficiently exhaustive. In the (2) \emph{reference} function, the same writing agent might request additional information from the user about the tone that is appropriate for the manuscript. That could be done via a simple question, but also via a more complex function, such as generating examples for the user to choose from, again, resulting in a sub-process that is in itself a control-loop. In the (3) \emph{output} function, the agent, while producing a sentence, might react to low confidence when writing a word by evoking a sub-process to double-check the spelling or to review whether an output is grammatically correct. Here again, the output function consists of at least one nested loop, which is utilized based on the system's confidence. 

In sum, characterizing AI systems as cybernetic control architectures with more and deeper nested loops preserves the conceptual core of automation research while accommodating the complexity of AI. However, this framing demands a well-defined description to ground research on: a task-centered account of human–AI interaction that specifies the joint loop in which humans and AI participate, but without reducing “human-in-the-loop” to a static checkpoint. In the forthcoming section we develop a loop-based model that (i) targets challenges introduced by increasing autonomy and adaptivity via interaction design rather than after-the-fact governance, (ii) formulates design questions for when humans and AI share the same regulatory loop at the task level, and (iii) characterizes human–AI interaction not only by outcomes but by the distinct sensing, deciding, and acting functions through which people engage with a task.

\section{A Shared Control Loop Involving Human and Machine}

Before presenting the model, we clarify the epistemic status of the IIP model in its current form. IIP is proposed as a conceptual framework and design heuristic, not as a fully validated predictive theory. Its primary function is twofold: first, to provide a structuring vocabulary that allows researchers and designers to describe human–AI interaction in terms of shared regulatory processes rather than isolated system features; second, to articulate integration qualities that can guide the selection of interface mechanisms and evaluation instruments. The model is designed to be compatible with, rather than competitive against, empirical theories of trust, situation awareness, or workload — it offers a scaffold onto which such constructs can be mapped and through which they can be related to concrete design decisions. Whether the integration qualities function as orthogonal dimensions, whether they predict user behavior beyond established constructs, and how they can be reliably measured are open empirical questions that we explicitly flag as directions for subsequent work. Readers should therefore engage with IIP as a theoretically grounded proposal for organizing the HAII design space, whose validation will require a dedicated empirical program.

The basis for the IIP model introduced in this section is the theoretical perspective of cybernetic loops of action-regulation, as discussed in Carver and Scheier’s work \cite{carverStructureBehavioralSelfRegulation2000}. An agent continuously compares a perceived state to a desired reference (goal or standard). Any discrepancy between the current and target state prompts action to reduce the gap, the consequences of the action update perception, and the comparison repeats. Consider a human-only example: a technician regulating the temperature of an industrial oven. They read the thermometer and observe the oven’s behavior (input function), hold a target temperature and tolerance in mind for the given situation (reference functions), and turn dials, open vents, or tweak fuel flow to reduce discrepancies between current and target state (output function). If the oven runs cold, they add heat; if it overshoots, they ease off.  Adapting this to our setting, the machine is assumed to run the very same loop. A controller (e.g., PID controller) senses current temperature (input function), compares it to a setpoint (reference function), and actuates heaters or fans (output functions), with nested routines for filtering noise, anticipating trends, and preventing overshoot.

To enable a fruitful discussion, we deliberately differentiate—function by function—how humans and machines realize each part of the loop and introduce separating terms. However, this differentiation does not oppose the idea that both humans and machines can be coupled within each of the described functions.  

\subsection{Input Function of Human and AI}
The \emph{input} function characterizes all the processes of human and machine agents to generate a situation model (i.e., the current state of the task and context \cite{committee_on_human-system_integration_research_topics_for_the_711th_human_performance_wing_of_the_air_force_research_laboratory_human-ai_2022}). Here, it is key to reflect humans' inherent capability to contextualize perceptions with meaning and their inability to communicate their sensory receptions without altering them. That is, following the definition of data and information from \cite{zins_conceptual_2007}, a human user usually does not share the raw signal (e.g., noise) with a machine but contributes information, e.g., the interpretation of a signal. Accordingly, while the AI system contributes \emph{machine data} (e.g., sensor readings, logs, database records, or textual corpora), the human user contributes \emph{human information} (e.g., tacit context, intentions, local constraints, or empirical interpretations of sensory stimulation). The combined input to the control loop of the task at hand is therefore the union of machine data and human information. If either side of the input function is incomplete, stale, or distorted, the loop begins from an impoverished basis, which degrades everything downstream. Consider a system that supports planning processes in emergency care: an algorithm might use individual health data to predict the resources needed for treatment (machine data), while the medical personnel know that a person not speaking the language will need additional resources (human information). Only when both forms of input are present does the loop start from an adequate situation model (cf. \cite{committee_on_human-system_integration_research_topics_for_the_711th_human_performance_wing_of_the_air_force_research_laboratory_human-ai_2022}). This example also demonstrates why the combination of both entities in the input function can be beneficial: they contribute different content, due to the nature of their information processing characteristics. In the given case, humans cannot stay up to date to current resources available and make predictions, but are able to use contextual information to enrich the prediction, because the medical device cannot easily assess language abilities. 

\subsection{Reference Function of Human and AI}
For the \emph{reference} function, the person and the system each maintain structures that enable them to identify divergence between the current state in a task and a desired state. On the one side, humans may possess a set of priorities, expectations, and methods for a given situation and point in time. These can be best captured by the concept of a "cognitive (reference) frame" as described in \cite{kleinDataFrameTheory2007}. 

We have deliberately chosen to label the human contribution in the reference function the \emph{Human cognitive frame}. This term denotes the person’s priorities, expectations, and procedures for what cues matter and which probes are worthwhile. Based on work in sensemaking, this account emphasizes that people may try to fit data to a frame \cite{kleinDataFrameTheory2007}). By doing so, the term signals the purposeful nature of human information processing and underlines the intrinsically driven adaptability within the reference function. 
In contrast, machines have a given purpose, either by training or design, and apply the corresponding methods instead of self-defining their purpose. Hence, the chosen term \emph{Machine algorithm} denotes the model or procedure (from rules to learned functions) that encodes objectives and priorities, such as a cost function trading off time, tolls, and safety proxies. In a shared loop, these two reference structures do not need to be identical, but they should avoid \emph{contradiction} on key aims and methods. For example, in community moderation, if a classifier implicitly rewards terse, punitive replies while the manager’s frame values mentoring newcomers, the pair will work at cross-purposes. If instead the algorithm is tuned to surface posts that invite constructive guidance, the human cognitive frame and the machine algorithm become mutually supportive.

\subsection{Output Function of Human and AI}
For the \emph{output} function, the key task is to focus on the possible resources human and AI can provide to address what resulted from the comparison of input and reference (i.e., the comparator in the control loop). For the IIP model, we assigned the term \emph{human capabilities} to describe the human contribution in the output function. These capabilities can include, e.g., forming hypotheses, making choices, and carrying out actions. In contrast, the system offers \emph{machine functionalities} that generate inferences or recommendations meant to support those actions or that incorporate functions that execute specific tasks (e.g., writing text or activating a mechanical device such as a thermostat). The point of sharing the loop is not to replace the human’s next step but to improve the possibility of successfully addressing the identified gap. For instance, a creative tool that merely returns a “style score” is less helpful than a tool that proposes two concrete layout alternatives aligned with the designer’s current hypotheses, because the latter directly enables selection and execution. In a household energy scenario, “Run the dishwasher at 22:00 to save an estimated €0.34 and reduce peak load” (machine functionality) is immediately actionable for the user’s capabilities, whereas “High tariff now” is accurate but far less usable. 

Taken together, these model components structure how the two agents - human and machine - act together in the frame of human-AI interaction, i.e., it provides a conceptualization of the quality of coupling two information processing entities within a shared task. 

\section{Integration Qualities based on Shared Human-AI Information Processing}

\begin{figure}[H]
    \centering
    \includegraphics[width=1\linewidth]{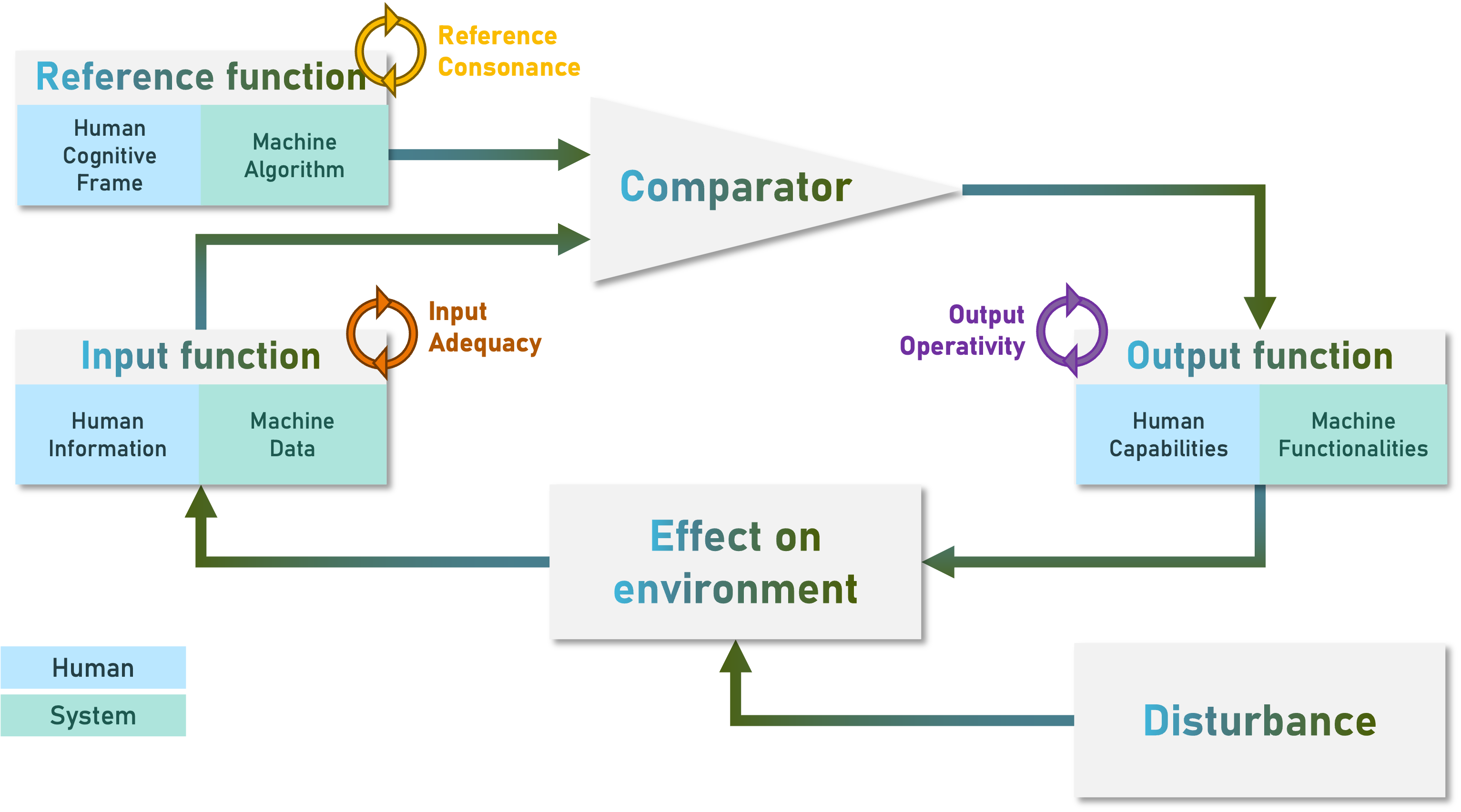}
    \caption{The Integrated Information Processing (IIP) model}
    \label{fig:IIP-full}
\end{figure}

As discussed before, the IIP model is designed to enable a structured discussion of human-AI interaction and corresponding design and evaluation. To this end, it is not sufficient to describe similarities and differences of human and AI information processing, but to define what constitutes an optimal coupling. In other words, a central contribution of the IIP model is the definition of integration qualities that constitute a successful coupling of human and AI, or, as we frame it within this model, to arrive at an integrated information processing and action regulation of humans and AI. The model, including all integration qualities, is depicted in Figure \ref{fig:IIP-full}.

Accordingly, we derived three information-processing qualities that can be used to understand and improve how well the coupling in human-AI interaction has been achieved: \emph{input adequacy}, \emph{reference consonance}, and \emph{output operativity}. For each integration quality (IQ), we first define the quality, then elaborate, and finally offer an example. In line with what we discussed earlier on nested loops, for each integration quality, agents also tend to hold a \emph{preferred level} for the quality and, when they sense a gap, they engage in a nested corrective loop: they detect the divergence, intervene (by adding or requesting something, or by adjusting settings or their own stance), and then re-assess until the perceived quality meets their preference. These inner loops sit inside the larger shared loop, and the possibility to invoke other entities when these preferred levels are not met is a key part of keeping effective coupling \cite{CarverScheier1982_ControlTheory, kleinDataFrameTheory2007, shneidermanHumanCenteredArtificialIntelligence2020a}.

\subsection{Input Adequacy}
The first integration quality concerns the input function. \textit{Input} on the one hand, includes information that a human actor can possess—e.g.,\ how much pain a person is currently experiencing or what events are going on in a city. All information a human is aware of and that is relevant for the given information-processing task (e.g.,\ making a decision) is considered part of human information. On the other hand, there is machine data that is available to the automated system (e.g., current capacity of the ICU in a hospital). To describe the extent to which the input function contributes to human-AI integration, we consider a term used in mathematics that describes the state of a model to be suitable for a given objective. Adequacy can be used to describe the state of a set of information that is sufficient to allow a conclusion \cite{tedeschiAssessmentAdequacyMathematical2006}. 

In light of the mathematical description of adequacy, input adequacy as integration quality is defined: ’\textbf{Input adequacy in integrated information processing exists when both entities have in total a sufficient amount of sufficiently correct information to process the task}’. Because the loop can only process what enters it, gaps or distortions on either side may produce errors or otherwise negative outcomes. As all integration qualities, adequacy has an objective dimension (what is actually present and what is sufficient information to solve a task perfectly) but also an experiential dimension (what the entities judge to be missing or unreliable). In a hiring scenario, for instance, an applicant-tracking system may parse résumés and scrape public profiles while a recruiter knows the impact of a candidate’s most recent project and the team’s unstated expectations. Input adequacy is high when the parsed artifacts and the recruiter’s information are both available and deemed to be valid; it is low if contract roles were missed by the parser or if the recruiter lacks the latest work sample. 

\begin{figure}[H]
    \centering
    \includegraphics[width=1\linewidth]{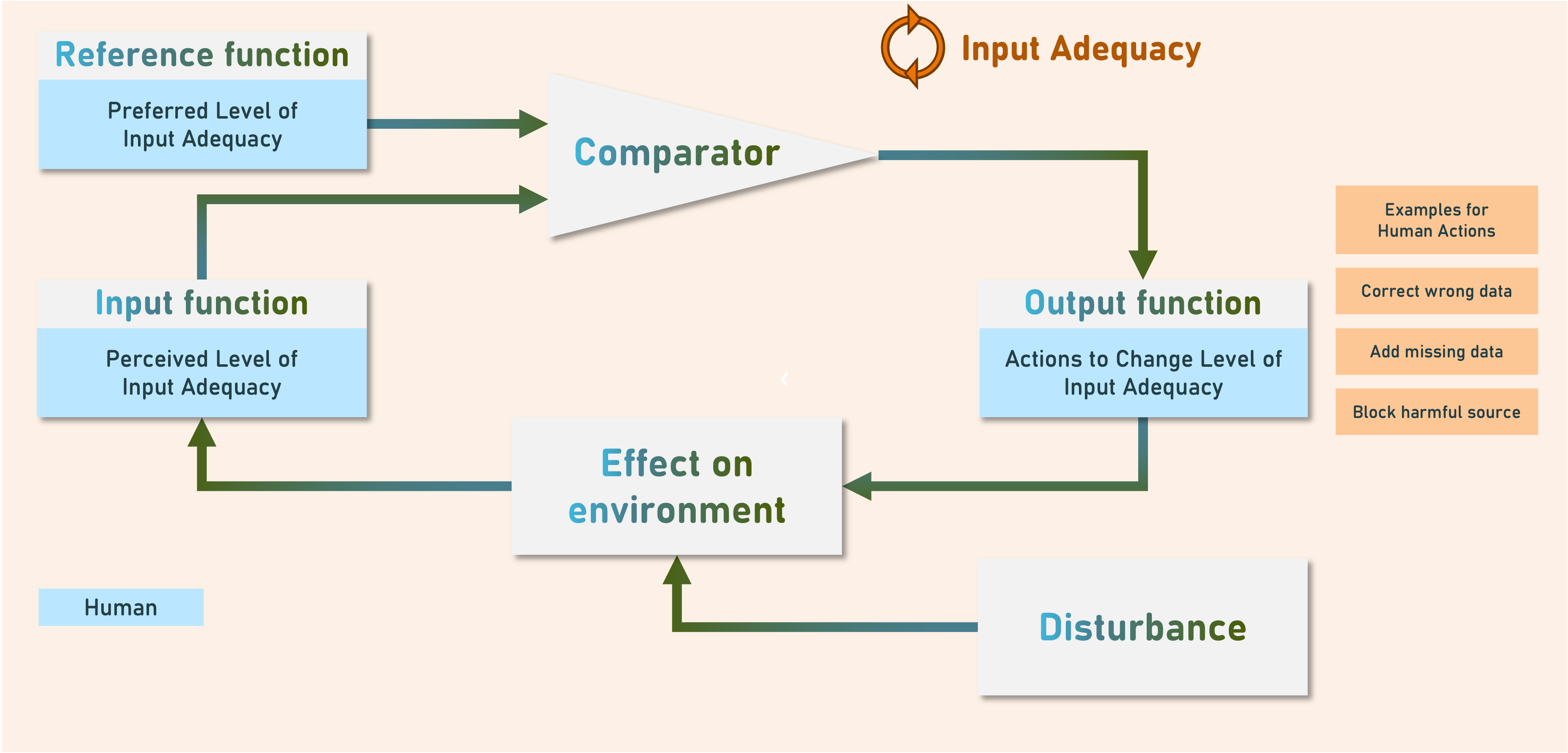}
    \caption{Depiction of the regulation of input adequacy of the user as a nested loop within input adequacy}
    \label{fig:IIP-IA}
\end{figure}

The above implies that both the human and the machine may conclude that input adequacy is insufficient, making further action necessary (also reflected in the idea of information asymmetry; see  \cite{hemmer_effect_2022}). Figure \ref{fig:IIP-IA} shows the nested loop of input adequacy, in which a comparison is made with the preferred level of input adequacy based on the available information. In the event of deviations, an output can be generated to increase input adequacy. For example, input adequacy could be below the human's preferred level because the system does not have the necessary information (e.g., the documents needed for a task). To correct this, the human could add this information, if possible.

\subsection{Reference Consonance}
The second integration quality concerns the reference function. In the context of automated information processing, the reference function describes more than a target value, as it does in simple control loops: it contains all processes and targets used to process the input. This may be, but is not limited to, specific target values (e.g., a maximum sentence length), a processing logic (e.g., if–then rules as to whether or not to use specific roads), or a neural network that represents a model. Hence, the reference function consists of objectives and procedures, while the input function contains data and information. For humans, it contains the cognitive frame, consisting, e.g., of goals, objectives, and methods.

Concepts comparable to an optimal coupling in the reference function exist in the literature on trust in AI, e.g., as goal alignment, by \cite{chiouTrustingAutomationDesigning2023c} or when looking at an AI system’s purpose \cite{benda2021purpose}. In the context of the IIP model, the term that best describes the dynamic of human and machine reference function is reference consonance. The term reference reflects the complexity of an algorithm or cognitive frame (as described above) better than the term goal, as a cognitive frame can include methods in addition to goals (see \cite{kleinDataFrameTheory2007}). The conceptualization as consonance instead of e.g., congruence or alignment avoids the connotation that the goals and processes of humans and machines must be identical. For example, target values can be represented differently in humans and machines (e.g., a duration of a trip or an arrival time), or different processes can be used for the same goal (e.g., a rule-based method and a deep neural network). Based on this description, it could be concluded that there is a lower congruence, although the integration quality would be fulfilled. The conceptualization as consonance makes it clear that even different descriptions of a reference do not necessarily stand in the way of integrated information processing. However, the different representations of the reference function in machine and human may represent an obstacle to integration if, for example, the human cannot interpret whether the algorithm and cognitive frame are in consonance or not.

Therefore, reference consonance is defined as follows: \textbf{Reference consonance in integrated information processing exists when the objectives and processes used to achieve those objectives by one entity do not contradict the objectives and processes of another entity}. Accordingly, perceived reference consonance describes humans' perception of the extent to which a machine’s algorithm does not contradict their cognitive frame. On the other hand, machines may calculate an estimation of the extent to which a human’s cognitive frame is free of contradictions to the machine’s algorithm.

\begin{figure}[H]
    \centering
    \includegraphics[width=1\linewidth]{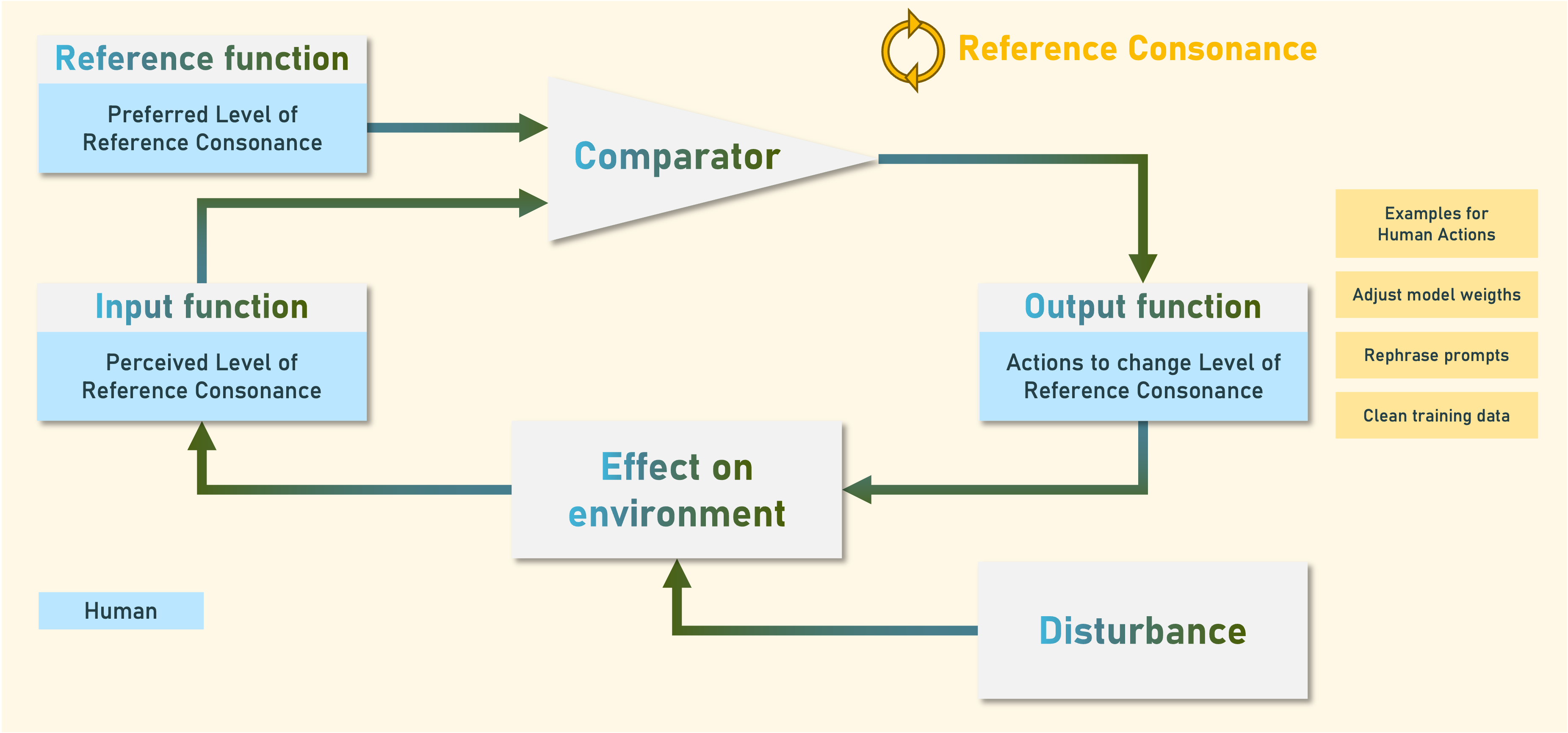}
    \caption{Depiction of the regulation of reference consonance of the user as a nested loop within reference consonance}
    \label{fig:IIP-RC}
\end{figure}

The given definition of reference consonance again implies that humans and machines can each recognize and respond to a level of reference consonance that diverges from their preferred level. Figure \ref{fig:IIP-RC} shows the nested loop of reference consonance, in which a comparison is made with the preferred level of reference consonance based on awareness of the other entity’s cognitive framework or algorithm. In the event of deviations, a reaction can be generated to correct the reference-consonance level. For example, reference consonance could be lower than the human’s preferred level because the system is biased towards undesirable outcomes (e.g.,\ an over-fitted system, lowering chances of female applicants). To correct this, the human could adapt the machine’s algorithm.

\subsection{Output Operativity}

The third integration quality concerns the \emph{output function}—what each entity does in the light of the comparison of input and reference (i.e., comparator result). In the IIP model, the human contributes \emph{human capabilities} (e.g., forming hypotheses, making choices, executing actions), while the system contributes \emph{machine functionalities} (e.g., generating inferences, recommendations, or directly executing tasks). However, we define an integration at this point to be a relevant metric of how both entities contribute to each other's ability to act effectively in relation to the task, reflecting research from joint human-agent activity and cooperation \cite{kleinTenChallengesMaking2004}. Accordingly, we define: \textbf{Output operativity in integrated information processing exists when human and machine are enabled to act in a way that is beneficial for the task at hand}. We deliberately use \emph{operativity} rather than, e.g., “actionability” or “applicability”: the former can be too narrowly action-centric (decisions are also part of the output function as described in \cite{carverStructureBehavioralSelfRegulation2000}), the latter too technical for the integration lens; still, insights from the actionability discourse help articulate what makes an output helpful for deciding among realistic alternatives. Operativity hinges on two complementary contributions in the loop: first, output should yield \emph{informational value for decisions}, especially by shifting the relative plausibility of current options and making differences between realistic alternatives salient (e.g., indicating that option~A reduces risk relative to option~B at similar cost); second, output should connect to \emph{feasible} next steps via available machine functionalities and human capabilities (e.g., lowering a room’s setpoint by $1^{\circ}\mathrm{C}$ now or revising the \emph{amount of credit required}), while avoiding proposals that cannot be executed (e.g., “change the applicant’s age”). 

\begin{figure}[H]
    \centering
    \includegraphics[width=1\linewidth]{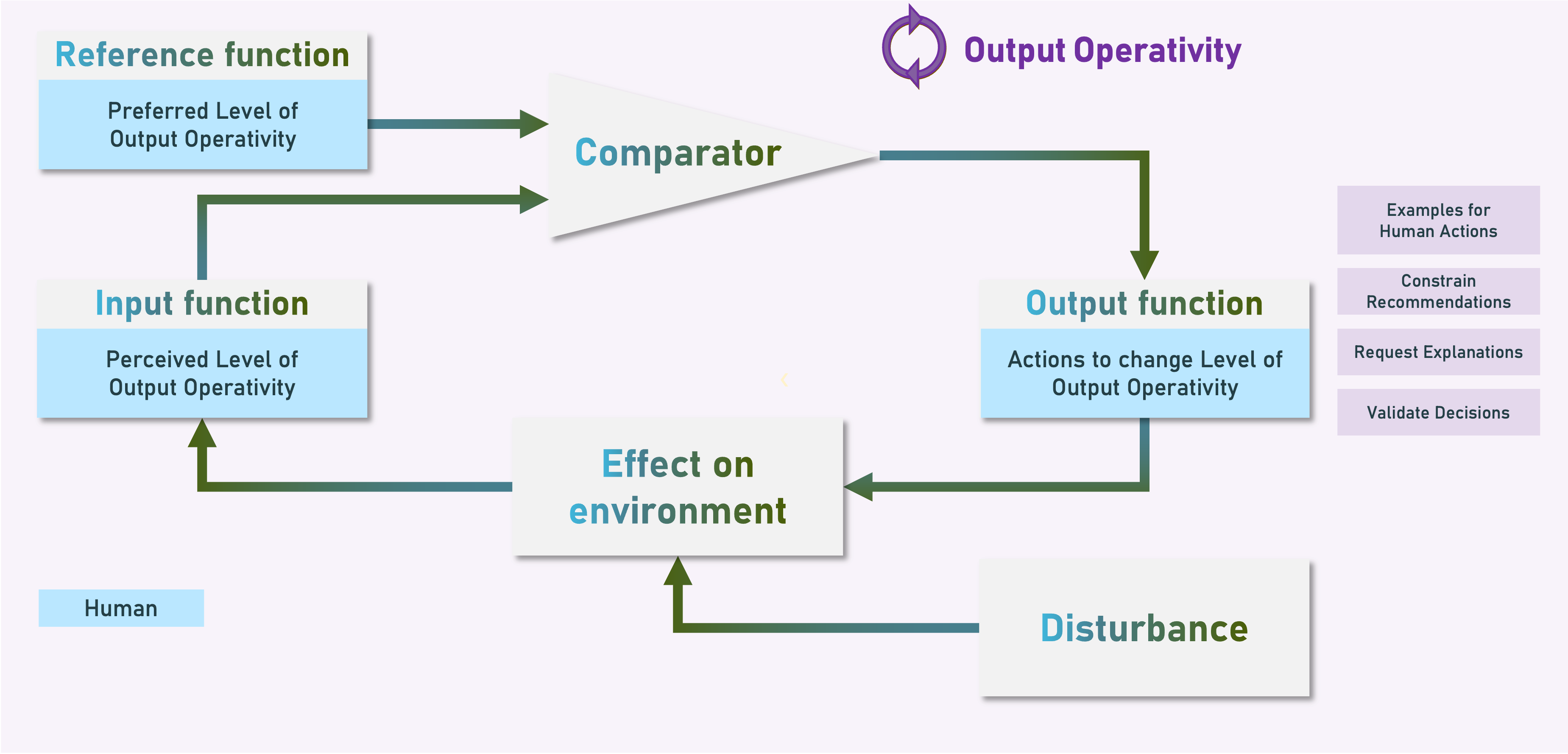}
    \caption{Depiction of the regulation of output operativity of the user as a nested loop within output operativity}
    \label{fig:IIP-OO}
\end{figure}

Two failure patterns illustrate low output operativity: the system produces a correct inference that nevertheless does not increase human capability (i.e., a conclusion without a comprehensible rationale or pointers to discriminating actions), or the human issues a command the system cannot fulfill (e.g., “change temperature in room~X” when no connected heater/actuator exists). In \emph{smart-home heating}, a thermostat that merely reports “high tariff now” is less operable than one that proposes feasible next steps and their expected effects—such as “lower living-room to $20^{\circ}\mathrm{C}$ for the next 2h; estimated savings and comfort impact shown”—and, where available, can execute the change or prepare a one-tap confirmation, aligning machine functionality (actuation and estimation) with human capability (choice and acceptance). In a \emph{writing assistant}, rather than returning a vague score (e.g., “style: 0.62”), an operable output offers discriminating options consistent with the user’s intent (e.g., two alternative introductions for a concise/neutral frame) and stands ready to insert the chosen paragraph and open references, making the suggestion both decision-diagnostic and immediately executable. As with the other integration qualities, agents compare perceived operativity against a preferred level and run an inner corrective loop when they detect a shortfall—specifically, when they realise that neither party can act or that they lack the basis to decide. Corrective actions include reformulating outputs to discriminate options, changing representation or granularity, surfacing feasible actions with confirmations, offering to carry out permissible steps, or requesting alternatives; the loop continues until perceived operativity reaches the preferred level. Because operativity describes whether outputs actually enable doing or deciding, it directly supports human control and oversight in the joint loop: systems should be designed to enable human capabilities—not merely to inform, but to make appropriate action possible. A depiction of a nested loop within output operativity for the human user is depicted in Fig \ref{fig:IIP-OO}.

\section{Application Fields of the IIP Model}

The key goal of the IIP model is to help structure and improve human-AI interactions through the lens of its three core components in joint action regulation and its proposed three integration qualities. Particularly, the model can (I) act as a structuring lens to analyze human-AI interactions, (II) ground interface modifications through theoretical motivation, and (III) help to structure the space of measures that try to indicate good coupling of humans and AI. 

\begin{tcolorbox}[title={IIP Core Constructs \& Examples},colback=gray!3,colframe=black,breakable,enhanced,left=6pt,right=6pt,top=6pt,bottom=6pt]

\textbf{Core constructs (definitions with one micro-example each).}
\begin{itemize}[leftmargin=1.15em]
  \item \textbf{Input adequacy:} Input adequacy in integrated information processing exists when both entities have in total a sufficient amount of sufficiently correct information to process the task.\\
  \textit{\footnotesize Example (micro-lever): show a compact “What I have” summary of recognized values with inline chips to edit/remove and a one-tap “Add missing field”.}

  \item \textbf{Reference consonance:} Reference consonance in integrated information processing exists when the objectives and processes used to achieve those objectives by one entity do not contradict the objectives and processes of another entity.\\
  \textit{\footnotesize Example (micro-lever): expose goals (e.g., \emph{min cost}, \emph{max quality}) with a  weight/threshold slider or presets, plus a “Why this?” rationale link.}

  \item \textbf{Output operativity:} Output operativity in integrated information processing exists when human and machine are enabled to act in a way that is beneficial for the task at hand.\\
  \textit{\footnotesize Example (micro-lever): present 2–4 clearly differentiated options, each with a primary \emph{Apply} action, a \emph{Preview/Insert} secondary, and instant \emph{Undo}.}
\end{itemize}

\medskip
\textbf{Designer takeaway.} Identify which quality currently limits integration (and drives undesired effects on user experience and behavior); apply the corresponding micro-lever and anticipate the directional change in that quality.

\end{tcolorbox}

\subsection{Structuring Empirical Patterns on Human-AI Interaction}

\subsubsection{Input Adequacy}
Illustrating the applicability of IIP on empirical data, input adequacy, e.g., explains a recurring pattern: when task-critical state information is missing from the joint information processing, human users try to complete and contextualize through their information. For example, with COVID-19 exposure-notification apps (e.g., \cite{ceron_covid-19_2021, CoronaWarnAppDocumentation2023}), users often tried to factor in situational facts—were we masked, was it indoors, how ventilated was the space?—because common risk score calculation were based on Bluetooth attenuation, exposure duration, and an infectiousness prior, not from mask or venue variables; interview studies with Canada’s COVID Alert further show people supplementing app notices with their own contextual reasoning when such variables are absent \cite{huang_users_2022}. 

As another example, in programming with AI code assistants, developers exhibit validation-and-repair behavior—opening tests, adding logs, checking docs, and rewriting suggestions—explicitly to supply the project/environment state the model cannot “see”, until the outputs become usable; controlled lab work with eye-tracking and IDE telemetry, and test-driven assistant workflows, document these compensatory actions \cite{mozannar_reading_2024, fakhoury_llm-based_2024}. More commonly, such mismatches show up in large software-adjacent communities where AI services lag behind fast-evolving norms or codebases. On Wikipedia, e.g., editors routinely actualize the model behind AI moderation tools: they catch misclassifications in real time (e.g., bad revert or quality scores), override them, and feed back labels or discussions that update evaluation sets and model behavior. Systems like ORES \cite{halfaker_ores_2020} expose predictions for patrollers while making data/model curation participatory, and newer tools like Wikibench let communities collaboratively build and debate evaluation datasets so the AI reflects current ground truth and community values \cite{kuo_wikibench_2024}.

Across these domains, the observable behavior—adding situational variables, instrumenting the task, updating shared state—may be described as users regulating input adequacy to restore a workable task representation. That is, the corrective actions (i.e., sub-loops) are an indicator of how well the system's input function aligns with users' situation modeling processes initially (i.e., input-related human-AI integration). 

\subsubsection{Reference Consonance}
Reference consonance captures how users act when they detect a mismatch between their own cognitive frame and the algorithm applied by the system—prompting them to correct, reweight, or restate constraints so the system does not contradict their pursuits. For example, in explanatory debugging, people use reasons shown by the system to locate where its internal criteria diverge from their own and then supply structured corrections (e.g., feature- or rule-level edits) so future behavior reflects their diagnostic logic; empirical evaluations show improved mental models and personalization once these corrections are incorporated, which is a positive effect of enabling correction of reference consonance actions \cite{kulesza_principles_2015}. 

As another example, in medical image retrieval for pathology, clinicians actively tune what counts as “similar”—adjusting which visual attributes the system should prioritize—because off-the-shelf relevance often conflicts with the pathologist’s case-specific notion of evidential value; refinement tools measurably increase diagnostic utility precisely by aligning the retrieval criterion with the expert’s reference frame \cite{cai_human-centered_2019, hegde_similar_2019}.
 
To illustrate that this corrective impulse extends beyond experts, user studies on explanations paired with feedback show that when people are given channels to react to model rationales, they both desire and use them to “fix” perceived flaws, and they hold the system accountable for updating accordingly—behavior that directly targets the reference \cite{smith-renner_no_2020}. 

In sum, behaviors such as demanding correction interfaces, redefining similarity between human-defined operating procedures and machine algorithm, or explicating goals to the system are best understood as users regulating reference consonance—aligning how the algorithm processed data with what they consider appropriate for the task.

\subsubsection{Output Operativity}
Output operativity captures how far an output actually enables the next step in the task, i.e., provides an operative benefit; when operativity is low, users may ask the system to make its recommendation actionable (e.g., by disambiguating alternatives, proposing concrete next tests, or giving viable recourse) or, if they have no control, may disengage. For example, in diagnostic support, clinicians explicitly seek differential-making help rather than post-hoc justifications: studies show that the type of explanation (e.g., local saliency vs. global prototypes) measurably changes physicians’ diagnostic performance, underscoring that explanations matter insofar as they help users discriminate among options and act \cite{xie_chexplain_2020}; when such guidance is missing, clinicians may negotiate with, delay, or ignore the advice instead of following it. 

As another example, in algorithmic advice settings, people judge systems by whether the suggested changes are feasible and actionable; user studies find that participants deem recourse “unreasonable” when it proposes impractical or immutable changes (e.g., systems suggesting a change of age or gender) and that their willingness to act depends on how recourse is framed and presented—some formats (e.g., simple feature guidance) are preferred or better understood than abstract counterfactuals \cite{singh_actionability_2025, ustun_actionable_2019}. 

In sum, behaviors such as asking for differential guidance, rejecting infeasible recourse, or abandoning tools that do not yield executable next actions are best read as users regulating output operativity: they either push the system towards actionability or, failing that, withdraw because the output does not enhance their capacity to carry the task forward.

\subsection{Designing for Human–AI Integration}

Beyond a more structured understanding of human behavior and experience, the IIP model can be used to specify design intent and predict how features of user interfaces as well as interaction strategies, should change the coupling between human and machine. In this view, each integration quality can act as guidance for designing user interfaces of intelligent systems. Below, we articulate and illustrate how recent examples from the field of HCAI can be described utilizing the IIP model.

\subsubsection{Input Adequacy}
In essence, the concept of input adequacy prompts interface and interaction design towards the question: is the exchange of information between system and user sufficient and do both have methods to detect and react to insufficient information?

Designs that surface the machine’s task state and create channels for correcting or adding data relevant to the task are connected to input adequacy. For example, in a study from Schrills \& Franke about automated insulin delivery, an information-disclosure explanation shows what the system currently “knows” (e.g., glucose level, insulin on board), thereby enabling users to judge whether the inputs suffice for the task \cite{schrillsHowUsersExperience2023a}; instructions that highlight information interdependence further helps users keep input adequacy high by prompting corrections when machine data are incomplete \cite{Schrills2025}. As an example from explainable conversational interfaces, displaying recognized information (e.g., location, detected intents) makes the system’s parsed inputs inspectable—even across modalities—so users can add or amend missing task variables \cite{schrillsExplainabilityCaseStudyConversational2021}. 
 
\subsubsection{Reference Consonance}
In addition, the concept of reference consonance provides a guiding principle for effective AI design (and research): do user and system have a model of each other's reference for a task, and are they able to influence, or correct, each other?

When design explicitly exposes what the system optimizes and offers levers to correct its criteria or procedures, it supports reference consonance. For example, XAI systems can present model-reasoning views (e.g., feature attributions, rule/rationale displays) and editable constraints (thresholds, rule toggles) that support users in reconciling algorithmic procedures with their cognitive frame (e.g., does the system use a similar structural model of what constitutes a "good" state as I do). As a concrete example, showing “why” alongside controls to change how the system weighs evidence lets users remove contradictions (e.g., exclude ethically problematic features) or adapt decision logic without changing the task’s facts. The predicted effect is that fewer resources are needed for correction, higher perceived alignment, and clearer accountability because the interface makes the reference function inspectable and adjustable, not merely legible. 
 
\subsubsection{Output Operativity}
Lastly, the concept of output operativity aligns with previous objectives from research on user acceptance and usage of automation, especially with usefulness: are users and systems able to address an identified gap, and can they produce inferences that enable such activities?

For example, systems in the field of evaluative AI \cite{miller_explainable_2023} bundle supporting and contradictory evidence so that users are enable to discriminate alternatives and decide what to do next; alternative research paths further recommends interactive explanations (e.g., what-if, scenario probes) or forcing functions that require users to commit to a preliminary judgment before seeing the AI, thereby reducing passive acceptance and increasing task engagement. This is well illustrated by an experiment on cognitive forcing, which showed that requiring users to form an initial judgment or verify key evidence reduces overreliance relative to a simple explanation condition—an operativity gain because the design compels a concrete next action in the workflow \cite{BucincaMalayaGajos2021}. Finally, further research positions implications for digital contact-tracing apps specifying diagnostic outputs as actionable guidance (e.g., which next behaviors to take), predicting higher perceived usefulness when outputs map to feasible actions. 

\subsection{Evaluation of Human–AI Interaction}
Finally, the IIP model can be utilized to structure the evaluation of systems and provide guidance for selecting and establishing optimal benchmark variables for good human-AI interaction. That is, quantitative metrics, scales, and indicators as well as qualitative probes, can be structured from the perspective of the three key integration qualities. Hence, the IIP could be useful to compare the results of different studies, although they did not use the exact same instrument or variable, but addressed the same integration quality. We will present examples for items from scales and how they can be connected to integration qualities. However, more often than not, we focus on specific items of these scales because the instruments are usually designed to capture broader and more holistic constructs. 

For \emph{input adequacy},  our first example is drawn from the System Causability Scale (SCS), a key scale, that is frequently used in Human-AI interaction \cite{holzingerMeasuringQualityExplanations2020a}. Here, participants are asked to respond to the following item: “\emph{I found that the data included all relevant known causal factors with sufficient precision and granularity},” which is a direct check of whether inputs capture the task state at the right resolution and therefore relate to input adequacy. Second, the Human–Computer Trust (HCT) scale \cite{madsenMeasuringHumanComputerTrust2000} covers the integration quality from the user’s vantage point with, “\emph{The system makes use of all the knowledge and information available to it to produce its solution to the problem}”, Third, Trust-in-Automation scale from Körber contain items that further address the system’s ability to correctly assess the situation, e.g., “\emph{The system is capable of interpreting situations correctly}” \cite{korber_theoretical_2019}. In cooperative teaming, Attig et al. include two the following two items: “\emph{the agent had access to all relevant information}” and “\emph{the agent shared relevant information with me},” both indexing whether the agent can \emph{ingest and surface} task state required for joint work \cite{attigMoreTaskPerformance2024}. In this case, the second item even refers to the system's ability to react to low levels of input adequacy by sharing relevant information.  

For \emph{reference consonance} first, SCS tests whether users need extra \emph{external} authorities to reconcile the explanation, e.g., “\emph{I did not need more references in the explanations: e.g., medical guidelines, regulations},” which indicates that the rationale already matches the user’s cognitive frame \cite{holzingerMeasuringQualityExplanations2020a}. Second, HCT targets the algorithmic side with, “\emph{The system uses appropriate methods to reach decisions},” a compact statement of method–criterion fit \cite{madsenMeasuringHumanComputerTrust2000}. Third, in a trust scale adaptation from Roesler et al. \cite{roesler_trust_2022}, “\emph{I am well informed how the system works}”, the item captures whether users have sufficient procedural knowledge to \emph{judge} fit and call out contradictions. In addition, cooperativity items such as “\emph{the agent knew enough about my goals to collaborate with me}” explicitly probe whether the agent has internalized the human’s objectives well enough for coordinated action \cite{attigMoreTaskPerformance2024}. Finally, actionability research treats contextual fit as relevant: “\emph{The information is socially appropriate},” which captures whether the content of information aligns with situational norms relevant to using it \cite{singh_actionability_2025}. 

For \emph{output operativity} SCS includes the controllability facet, “\emph{I could change the level of detail on demand},” which lets users tailor information to the action at hand \cite{holzingerMeasuringQualityExplanations2020a}. HCT offers a plain operativity probe: “\emph{The system always provides the advice I require to make my decision}” \cite{madsenMeasuringHumanComputerTrust2000}. In applied trust adaptations, “\emph{I find that the system supports my work}”, the given question addresses whether outputs translate into practical support \cite{roesler_trust_2022}. Thematically most fitting, work on actionability provides items that measure operativity very direct: “\emph{The information allows me to break down any action into explicit steps}” and “\emph{The information helps me understand the reason(s) for the decision},” the latter linking justification to a \emph{usable} course of action \cite{singh_actionability_2025}.

All in all, it can be seen that frequently used measures in the human-AI interaction literature trace aspects of the cybernetic integration qualities, which gives some indication that the qualities are relevant for the evaluation of intelligent systems. Hence, when selecting or developing any new type of benchmark values, or simply creating an evaluation strategy in a single research or development project, it can be highly beneficial to structure the included instruments from the perspective of the integration qualities.

\section{Discussion}
Having derived and explicated the application fields of the IIP model, the objective of the following section is to discuss the key paths and needs for the further development of the IIP model addressing its current limitations: (i) structuring optimal automation patterns within human-AI interaction, (ii) a further conceptualization of human-AI integration, and (iii) how to advance instruments for evaluation of human-centered AI from a cybernetics perspective.

\subsection{From Levels of Automation to Integration Qualities}

The IIP model is intended as a concise lens for structuring human-AI integration. It deliberately shifts the central question in automation design in human-AI systems away from choosing a particular \emph{degree of automation} toward deciding \emph{what we ultimately target when we design the automation pattern}. Thus, while the IIP model addresses automation, it tries to move forward from the classic stage- or level-based accounts of automation by rather emphasizing the specific \emph{qualities} of the coupling itself (i.e., compared to placement of design decisions on a scale of automation intensity). That is, it moves away from putting an engineering view first (e.g., defining system functions) to putting a human view first (e.g., the psychological experiences and action regulation patterns one wants to achieve). 

However, while the IIP model may already be applicable as a conceptual model or design model, as a psychological framework, it should also try to enable the construction of \emph{strict hypotheses}. In order to do so, it provides the scaffolding from which system-specific hypotheses can be derived: e.g., which concrete design changes are expected to cause which shifts in the integration qualities, and how integration qualities can impact user experience and action regulation. Achieving an examination of model-based hypotheses requires tighter ties between observed behavior and available actions in the interface. Some of this linkage is already conceptually captured by the nested corrective loops in the model. However, future empirical work is needed to deepen the link between design decisions and behavior based on IIP so that changes in interface and interaction can be traced and, in turn, designers can deliberately create changes in integration. Such empirical advances and grounding of the model would enable testing and defining the detailed mechanisms of the integration qualities to enable deriving more specific predictions of user behavior and experience.

\subsection{Conceptualizing Integration}

A second key question of the IIP model is to conceptualize how frequently used constructs of human-automation (and human-AI) interaction relate to integration. Trust (or perceived trustworthiness) as one frequently used automation-related variable of user experience, i.e., can most likely be understood as influenced through \emph{all} integration qualities rather than through a single component of the cybernetic loop of joint action regulation. This can also be demonstrated by analyzing typical instruments to assess trust (see Section Model Application). In turn, the IIP model does not provide a clear explanation of how, e.g., low levels of trust are connected with integration qualities. We argue that this is due to the abstract and sometimes not-overlapping definition of trust (e.g., \cite{hoff_trust_2015} and also because trust is possibly not as closely connected to behavior as assumed in previous research (see \cite{boltonTrustNotVirtue2022}). 

Moreover, from a more foundational perspective, it is interesting to note that many benchmark concepts in human-centered AI were originally adapted from human-human interaction (e.g., interpersonal trust carried over to trust in automation, teaming and cooperation mapped to human--AI settings). Hence, they carry a highly psychological, human-focused notion in their conceptualization. While this approach has many advantages, such as a typically rich existing research literature in basic and applied psychology and a high face validity resulting in the construct being easy to communicate in scientific discourse, it also has its structural problems because these constructs were not explicitly developed to target the specific dynamics in human-machine interaction and they represent tendencies to anthropomorphize technology.

Oppositely, other benchmarks have been explicitly developed from the targets defined in technical system feature conceptualizations-i.e., can be described as a kind of target responses (e.g., explanations in XAI should subjectively satisfy users = explanation satisfaction as benchmark). Hence, they carry a strong engineering feature-focused notion without a strong connection to processes of human psychology. 
While these measures provide helpful guidance in design iterations and evaluations, they were not intended to have a strong theoretical embedding with information processing, but are on a meta-level of the interaction. The IIP model, in contrast, is designed to emphasize the \emph{psychological structure of the interaction} with systems that exhibit a high degree of agency. 

At the same time, IIP is not offered as a substitute for existing constructs or measures. It can help to \emph{structure evaluation}, but empirical research is needed (e.g., studies focused on mapping the construct relations with established research strategies for investigating construct validity) before the exact relations with established constructs can be defined. Experienced uncertainty in users' action regulation is one illustrative example here: while the IIP model may allow for understanding where uncertainty can arise in the interaction loop, there is a need for empirical proof of these assumptions before it can be utilized as guidance in a design process.

\subsection{Advancing Conceptualization of Integration}

Constructs are important conceptual tools to structure human experiences and behavior, yet they only contribute in practical processes when their relevance can be measured in a meaningful way. For measurement and evaluation of integration qualities, the next step is a structural analysis of the core cognitive and behavioral strategies that users employ to compensate for deficits in each integration quality (e.g., how users correct for a lack of input adequacy). Such analyses, grounded in empirical studies, can provide indicators (e.g., items or behavioural patterns) of suboptimal integration and help identify key facets or factors of integration quality without collapsing them into general satisfaction-type responses.

IIP can, in principle, be extended to settings with multiple entities on both sides, including agentic AI systems operating together with multiple human actors. However, doing so raises questions of hierarchy and role that can make the loop model fuzzier. Moreover, spaces populated by multiple systems and agents are only now beginning to be addressed by empirical research \cite{zheng_ux_2022, dahiya_survey_2023, zhang_see_2024}; they did not directly shape the model presented here, but theory should guide this work as it accumulates.

Finally, we consider whether integration qualities are design goals. Integration can indeed function as a design goal, yet there are valid reasons why integration may be hindered or only partially achievable in practice. Some levers improve integration (in particular, explanatory and XAI mechanisms that support interpretation, correction, and action), but they do not relieve the need to decide \emph{how far} we want integration to go. In analogy to levels of automation, there may ultimately be meaningful levels of integration. We treat this as an outlook rather than a present commitment: it is a direction for future refinement once more applications and empirical evidence are available.

\subsection{Relation to Existing Frameworks}

Given the long tradition of modeling human–machine coupling, a new framework must justify its existence by showing what it addresses that established models do not. We therefore position the IIP model in relation to four influential frameworks that share conceptual territory with our approach: Parasuraman, Sheridan, and Wickens' model of types and levels of automation, Klein and colleagues' Joint Cognitive Systems and their "Ten Challenges" for human–agent teaming, Chiou and Lee's framework of responsivity and resilience in automation trust, and Shneiderman's two-dimensional framework of human-centered AI. Rather than claiming that IIP supersedes these frameworks, we argue that it occupies a distinct position in the conceptual space they collectively span — and that this position becomes increasingly relevant as AI systems exhibit higher degrees of agency.

Parasuraman, Sheridan, and Wickens (2000) introduced a model that decomposes automation along four information-processing stages (information acquisition, analysis, decision, and action) and ten levels of automation per stage. This model has been enormously productive for analyzing function allocation and has shaped decades of empirical work on automation effects. Its orientation, however, is primarily engineering-centered: it helps designers decide what and how much to automate, but it treats the human largely as the non-automated remainder of the loop. The psychological processes by which users regulate their engagement with the system are addressed as consequences of automation choices (e.g., complacency, skill degradation) rather than as constituents of the loop itself. The IIP model complements this view by modeling both human and machine through the same cybernetic vocabulary, and by asking not how functions should be allocated but how the resulting coupling can be characterized in terms of integration qualities. Where Parasuraman et al.'s framework answers "what should the machine do?", IIP answers "what should the coupling between human and machine achieve?"

Klein, Woods, Bradshaw, Hoffman, and Feltovich (2004) formulated ten challenges that automated systems must meet to function as team players in joint human-agent activity, including mutual predictability, directability, shared situation awareness, and coordination costs. This account is closest to IIP in spirit: it is explicitly task-centered, treats joint activity as the unit of analysis, and emphasizes the bidirectional nature of coupling. However, the "Ten Challenges" are formulated as requirements — a checklist of properties a system should exhibit — without a unifying structural account of where in the interaction loop these properties take effect. IIP can be read as providing such a structural account: the ten challenges can largely be mapped onto the three functions (input, reference, output) and their corresponding integration qualities. For example, mutual predictability is a property primarily located in the reference function and assessed through reference consonance, while directability relates to output operativity. In this sense, IIP does not replace Klein et al.'s account but provides the mechanistic scaffolding that makes their requirements addressable through specific design interventions.

Chiou and Lee (2023) recently proposed responsivity and resilience as design principles for trustworthy automation, emphasizing that trust calibration depends on how systems respond to changing goals and conditions. Their framework shares with IIP the move away from static properties (such as "trustworthiness") toward dynamic characterizations of the human–machine relationship. The key distinction is one of focus: Chiou and Lee center their account on trust as the target construct, whereas IIP treats trust as one of several user experiences that emerge from — but are not identical to — the underlying integration qualities. IIP thus offers a layer beneath trust-centered accounts: it models the regulatory coupling that, among other things, shapes whether trust calibration is possible. This positioning also clarifies why IIP is not meant to replace trust research but to provide structural anchors for it.

Shneiderman's (2020) human-centered AI framework advocates for systems that are simultaneously highly automated and highly controllable, rejecting the tradeoff assumed in classical levels-of-automation accounts. IIP is compatible with this stance and can be read as offering a process-level account of what Shneiderman describes at a system-level: output operativity in particular operationalizes the idea that high automation and high human control are not mutually exclusive but depend on how outputs are designed. Where Shneiderman's framework is programmatic, IIP is mechanistic; the two operate at different levels of abstraction and can be used jointly.

Taken together, the distinctive contribution of the IIP model is not the introduction of new primitives but the combination of three properties that, to our knowledge, no existing framework offers jointly: (1) a unified cybernetic vocabulary that applies symmetrically to human and machine, (2) a task-centered structural decomposition into three functions rather than a stage- or level-based decomposition of automation, and (3) the explicit specification of integration qualities as design and evaluation targets, together with nested corrective loops that describe how these qualities are regulated during interaction. It is this combination — not any single element — that we argue makes IIP a useful addition to the conceptual toolkit of HAII research.

\section{Conclusion} 

This paper proposed the Integrated Information Processing (IIP) model as a compact, task-centred account of human–AI integration built on cybernetic control loops shared by human and machine. By using a unified vocabulary for input, reference, and output functions and by introducing three integration qualities—input adequacy, reference consonance, and output operativity—we connected psychological mechanisms of action regulation with concrete interface and system design choices. This perspective preserves continuity with established automation research while accommodating the higher autonomy and adaptivity of contemporary AI systems, and it yields practical criteria for analysis, design, and evaluation of HAII systems. 

Beyond offering a structuring lens, the IIP model motivates specific hypotheses: interfaces that surface and let users correct task state should raise input adequacy; controls that expose and tune optimization criteria should improve reference consonance; and outputs that discriminate among realistic alternatives and connect to feasible next steps should increase output operativity. Future work should empirically examine these predicted links, develop reliable indicators for each quality (including users’ corrective sub-loops), and extend the account to multi-agent configurations and potential \textit{levels of integration}. Taken together, IIP reframes human-centredness as the quality of the coupling itself—how well humans and AI co-regulate a shared task—and thus offers a principled foundation for designing intelligible, controllable, and effective interactive AI.

\makeatletter
\if@ACM@anonymous
  \section*{Disclosure of AI Assistance}
\else
  \section*{Acknowledgments}
\fi
\makeatother
We used OpenAI's ChatGPT (GPT-5 Thinking; version from October 2025; \url{https://chatgpt.com}, last accessed October 10, 2025) to improve readability and refine the English language. We reviewed and edited all outputs and take full responsibility for the final content.

\bibliographystyle{ACM-Reference-Format}
\bibliography{bibliography}
\end{document}